\documentstyle[aps,epsf,epsfig]{revtex}
\def \be   {\begin{equation}}
\def \ee   {\end{equation}}
\def \l {\label}
\begin{document}
\input epsf
\baselineskip=25pt
\title{FINITE DISCRETE ELECTROMAGNETIC FIELD THEORY}
\author{Manoelito M de Souza, Jair V. Costa, Adriano S. Pedra}
\address{Universidade Federal do Esp\'{\i}rito Santo - Departamento de
F\'{\i}sica\\29065.900 -Vit\'oria-ES-Brasil}
\date{\today}
\maketitle
\begin{abstract}
\noindent  The classical electromagnetic field of a spinless point electron is described in a formalism with extended causality by discrete finite  point-vector fields with discrete and localized point interactions.  These fields are taken as a classical representation of photons, ``classical photons". They are all transversal photons; there are no scalar nor longitudinal photons and the Lorentz gauge condition is automatically satisfied. The angular distribution of emitted photons reproduces the directions of maximum emission of the standard formalism. The Maxwell formalism in terms of continuous and distributed fields is retrieved by the smearing of these discrete fields over the light-cone, and in this process scalar and longitudinal photons are necessarily created and added. Divergences and singularities are by-products of this averaging process. The discrete and the continuous formalisms are not equivalent. The discrete one is superior for not having divergencies, singularities, unphysical degrees of freedom, for describing processes of creation and annihilation of particles instead of advanced solutions, for having a natural explanation for the photon, and for generating the continuous formalism as an effective one. It enlightens the meaning and the origin of the non-physical photons in the standard formalism. The standard theory based on average continuous fields is more convenient and appropriate for dealing with a large number of charges and for relatively large distances, but for few charges or for the field configuration in a charge close neighbourhood the discrete field description is mandatorily required for avoiding inconsistencies.
\end{abstract}
\begin{center}
\end{center}
\section{Introduction}
\noindent The concept of extended causality \cite{hep-th/9610028,hep-th/9610145,hep-th/9708066,hep-th/9708096} allows a formalism for field theory in terms of discrete point-like fields with localized point-interactions and free of singularities and divergences. One can see a classical field theory in terms of these discrete point-like fields as a kind of pre-quantum field theory. It is relevant for pointing the way to a finite description of fundamental fields in terms of point-like objects, a simpler and reliable possible alternative to string motivated approaches to a finite field theory. Its relevance increases with the anticipated information that this formalism has also successfully been applied \cite{gr-qc/9801040} for describing the gravitational field of General Relativity in terms of finite point-like fields, although this will not be discussed here. In this note we write Classical Electrodynamics in the context of extended causality, and we describe the discrete field of a classical spinless point electron.\\
For completeness, we start in Section II with a brief review from references \cite {hep-th/9610028,hep-th/9610145,hep-th/9708066,hep-th/9708096} about extended causality and its applications to classical field theory. We discuss the meaning of the fields and of their equations in this formalism, the Green's functions, their interpretation in terms of creation and annihilation of particles, and how the standard formalism is retrieved in terms of lightcone-averaged fields. Then we discuss the role of virtual (unphysical) photons. There is no virtual photons in a formalism with extended causality but they necessarily must be included for retrieving the standard formalism.\\ In Section III we show the structure of the radiation field of a spinless point electron. Section V analyzes the photon angular distribution. Finally, in Section V, we study the photon energy-momentum and we retrieve the Larmor Theorem.

\section{Extended causality}

\noindent In standard field theory formalism the evolution of a free field is constrained by
\be
\label{1}
\Delta\tau^2=-\Delta x^{2},
\ee 

where $\Delta\tau$ is the change of proper time associated to the four-vector  $\Delta x$  in a Minkowski spacetime of metric $\eta=diag.(1,1,1,-1)$. In our notation we omit the spacetime indices when this does not compromise the text comprehension. So, $x$ stands for $x^{\mu},$ $\; \partial$ for $\partial_{\mu}$, and $A(x,\tau)$ for a vector field $A^{\mu}(x,\tau)$, for example. Geometrically (\ref{1}) is the definition of a three-dimensional double cone; $\Delta x$ is the four-vector separation between a generic event $x^{\mu}\equiv({\vec x},t)$ on the cone and the cone vertex. See the Figure 1. The cone-aperture angle $\theta$ is given by
\be
\tan\theta=\frac{|\Delta {\vec x}|}{|\Delta t|},\qquad c=1.
\ee
A change of the supporting cone corresponds to a change of speed of propagation and is an indication of interaction.
Special Relativity restricts $\theta$ to the range $0\le\theta\le\frac{\pi}{4},$ which corresponds to a restriction on $\Delta\tau:$
\be
0\le|\Delta\tau|\le|\Delta t|.
\ee
\parbox[]{7.5cm}{
\begin{figure}
\hspace{-5.0cm}
\epsfxsize=400pt
\epsfbox{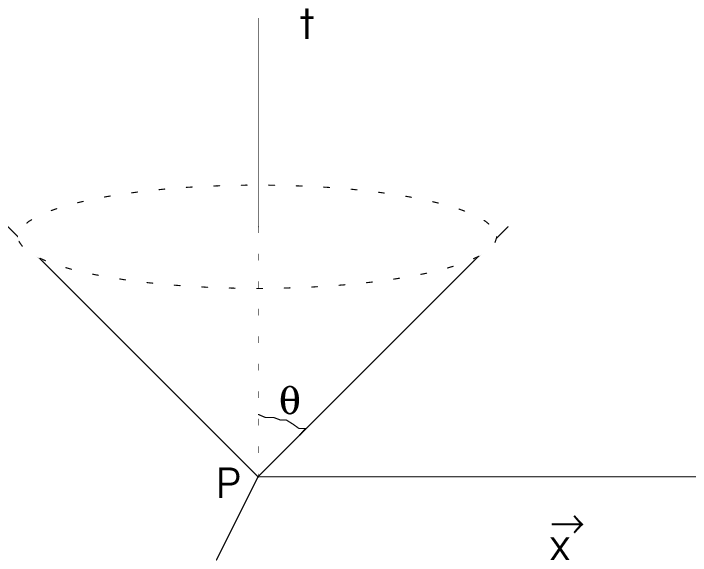}
\vglue-5cm
\end{figure}
\vglue-8cm%
}
{}\\
\mbox{}
\hfill
\hspace{5.0cm}
\parbox[]{7.5cm}{\vglue-5cm Fig. 1.
The relation $\Delta\tau^2=-\Delta x^{2},$ a causality constraint, is seen as a restriction of access to regions of spacetime. It defines a three-dimension cone which is the spacetime available to a point physical object at the cone vertex.}\\ \mbox{}
\vglue-1cm

This conic hypersurface is the support manifold for the field definition: a free field cannot be inside nor outside but only on the cone. Eq. (\ref{1}) is, therefore, a constraint on a physical object, a particle or a field, to keep it on the cone. This is local causality. The concept of extended causality corresponds to a more restrictive constraint; it requires that (\ref{1}) be simultaneously applied to an $x$ neighboring event $x+dx:\qquad (\Delta\tau+d\tau)^2=-(\Delta x+dx)^{2}.$  With $\Delta\tau^2=-\Delta x^{2}$ and $d\tau^2=-d x^{2}$ it is reduced to
\be
\Delta x.dx+\Delta\tau d\tau=0.
\ee
This is equivalent to the imposition of a second constraint, besides the first one (\ref{1}):
\be
\l{f}
d\tau+  f.d x=0.
\ee
$f$ is a constant spacelike $(f^{2}=-1$) four-vector tangent to the cone, a cone generator, and it is defined\footnote{Or, $\Delta\tau+f.\Delta x=0$ with $f\equiv\frac{d x}{d\tau}$, defining a constraint between $x$ and $\tau$. They are equivalent, for a free field.} by $f^{\mu}=\frac{\Delta x^{\mu}}{\Delta\tau},$ if $\Delta\tau\ne0;$ it is lightlike $(f^{2}=0$) in the case limit when $\Delta\tau=0$. \\
The equation (\ref{f}) can be obtained from direct differentiation of (\ref{1}), and geometrically it defines a hyperplane tangent to the cone (\ref{1}). We have from (\ref{f}) that $$f_{\mu}=-\frac{\partial\tau}{\partial x^{\mu}},$$ when $\tau$ is seen as a function of x, a solution of (\ref{1}): $\tau=\tau_{0}\pm\sqrt{-(\Delta x)^{2}}$.  For $d\tau=0,\;$ $f_{\mu}$ is orthogonal to the hyperplane (\ref{f}), but it is also a generator, at the lightcone vertex.\\ Imposing in field theory the two constraints, (\ref{1}) and (\ref{f}), instead of just (\ref{1}), as it is usually done, corresponds to knowing the initial position and velocity in point-particle dynamics.

\noindent Together, the constraints (\ref{1}), as $d\tau^2=-d x^{2}$, and (\ref{f}) are equivalent to the single condition $d x^2+(f.d x)^2=0$, or
\be
\l{lambda}
d x.\Lambda^{f}.d x=0,
\ee
with $\Lambda^{f}_{\mu\nu}=\eta_{\mu\nu}+f_{\mu}f_{\nu},$ ($f_{\mu}=\eta_{\mu\nu}f^{\nu}$),  which is a projector orthogonal to  $f^{\mu}$,  $  f.\Lambda.f=0.$ Therefore the constraint (\ref{lambda})   allows only the displacements $d x^{\mu}$ that are parallel to $f^{\mu}.$  A fixed four-vector $f$ at a point represents a fibre in the spacetime, a straight line tangent to $f^{\mu}$, the $f$-generator of a local cone (\ref{1}).
For a massless field (\ref{1}) defines a lightcone, (\ref{f}) defines a hyperplane tangent to this lightcone, and $f^{\mu}$ is the lightcone generator tangent to this hyperplane.

One can summarise it by saying that while the local causality restricts the available space-time of a free physical object to a conic three-dimensional hypersurface, the extended causality restricts it to just a straight line, a cone generator.
\begin{center}
Fields and field equations
\end{center}

\noindent $A_{f}(x)$ is a $f$-field, that is, a field defined on a fibre  $f$. It is distinct of the field $A(x)$ of the standard formalism, which is defined on the cone. $A_{f}(x)$ may be seen as the restriction of $A(x)$ to a fibre $f$ 
\be
\l{Af}
A(x,\tau)_{f}=A(x,\tau){\Big |} _{dx.\Lambda^{f}.dx=0}
\ee
It is a point-like field, the intersection of the wave-front $A(x,\tau)$ with the fibre $f$. This definition (\ref{Af}) would not make any sense if the discrete and localized character of $A_{f}$ could not be sustained during its time evolution governed by its wave equation. But it is remarkable that, as we will see, its point-field character remains as it propagates along its fibre $f$. As its electric field and its energy-momentum content remains invariant in time, $A_{f}$ can be seen as a linearly polarized ``classical photon".
Conversely, we have the following relation 
\be
\label{ss}
A(x,\tau)=\frac{1}{2\pi}\int d^{4}f\;\delta^2(f^2)\;A_{f}(x,\tau),
\ee
where $d^{4}f=df_{4}\;|{\vec f}|^{2}\;d|{\vec f}|\;d^{2}\Omega_{f}$. In the source instantaneous rest-frame, at the emission time (see eq. (\ref{f4})), we have for the emitted ($f^{4}=|{\vec f}|$) field,
\be
\label{s}
A(x,\tau)=\frac{1}{4\pi}\int d^{2}\Omega_{f}A_{f}(x,\tau),
\ee
where the integral represents the sum over all $f$ directions on the cone (\ref{1}). $4\pi$ is a normalization factor, $4\pi=\int d^{2}\Omega_{f}.$ The physical interpretation associates $A_{f}(x)$, a point-perturbation propagating along the lightcone generator $f,$ with a physical photon - we call it a classical  photon - and $A(x)$, the standard continuous field, to the effect of the classical photon smeared on the lightcone spacetime. It is worthwhile to anticipate here a remark about the physical distinction between $A_{f}(x,\tau)$ and $A(x,\tau)$. They do not represent  equivalent physical descriptions. $A_{f}(x,\tau)$  corresponds to a single real physical photon with $f$ being its four-vector velocity and with, as we will see, transverse electromagnetic fields while $A(x,\tau),$  due to the smearing process (\ref{s}), corresponds to a continuous distribution of fictitious unphysical photons with longitudinal electromagnetic field. This explains all the unexpected difficulties we have on quantizing the Maxwell field $A(x,\tau)$. Another remarkable distinction is that $A_{f}(x,\tau)$ is a finite pointwise field while $A(x,\tau)$ has a singularity introduced by the smearing process (\ref{s}). Figure 2 shows the relationship between the fields $A_{f}$ and $A$ for a process involving the emission of a single physical photon $A_{f}(x)$; $A$ here is its space average.

\noindent The derivatives of $A_{f}(x),$ allowed by the constraint (\ref{lambda}), are the directional derivatives along $f,$ which with the use of (\ref{f}) or of (\ref{lambda}) we write as
\be
\label{fd}
\partial_{\mu}A_{f}=(\frac{\partial }{\partial x^{\mu}}+\frac{\partial \tau}{\partial x^{\mu}}\frac{\partial}{\partial \tau})A(x,\tau){\Big |} _{dx.\Lambda^{f}.dx=0}={\Big(}\frac{\partial }{\partial x^{\mu}}-f_{\mu}\frac{\partial}{\partial \tau}{\Big)}A_{f}\equiv\nabla_{\mu} A_{f}.
\ee
With $\nabla$ replacing $\partial$ for taking care of the constraint (\ref{lambda}), the propertime $\tau$ is treated as a fifth independent  coordinate. 

\parbox[]{7.5cm}{
\begin{figure}
\vglue-6cm
\epsfxsize=400pt
\epsfbox{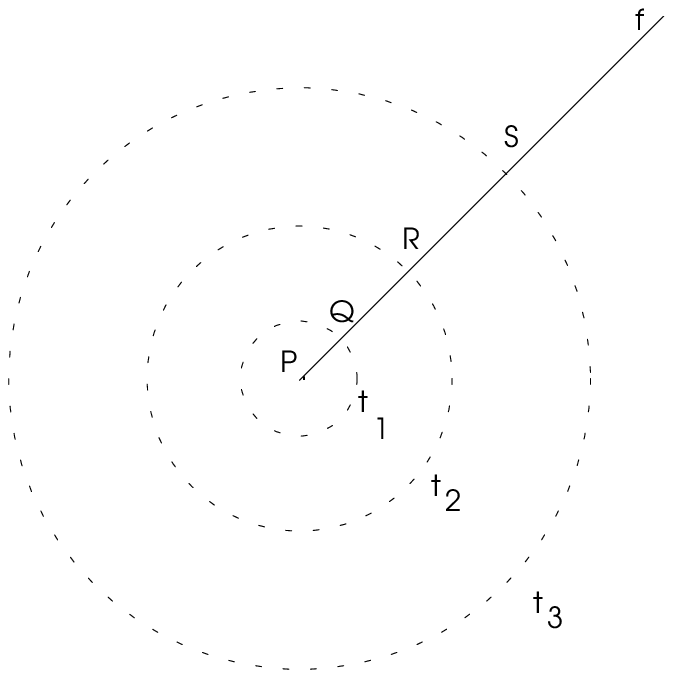}
\vglue-10cm
\end{figure}
\vglue-0cm}\\
\mbox{}
\hfill
\hspace{5.0cm}
\parbox[]{7.5cm}{\vglue-1cm Fig. 2.
The relationship between the fields $A_{f}$ and $A$. The three doted circles represent, at three instants of time, the field $A$ as an spherically symmetric signal emitted by a charge at the point P. The straight line PQRS\dots is the fibre $f,$ a lightcone generator tangent to $f^{\mu}.$ The points Q, R, and S are a classical photon $A_{f}$ at three instants of time.}\\ \mbox{}

\noindent The field equation for a massless field defined on a lightcone generator $f$ is, consequently,
\be
\label{wef}
\eta^{\mu\nu}\nabla_{\mu}\nabla_{\nu}A_{f}(x,\tau)=J(x,\tau),
\ee
or, explicitly
\be
\label{wef'}
(\eta^{\mu\nu}\partial_{\mu}\partial_{\nu}-2f^{\mu}\partial_{\mu}\partial_{\tau})A_{f}(x,\tau)=J(x,\tau),
\ee
as $f^{2}=0$. $J$ is its source four-vector current.\\ 
An integration over the $f$ degrees of freedom in (\ref{wef}) reproduces, with the use of (\ref{s}), the usual wave equation of the standard formalism, \be
\l{10'}
\eta^{\mu\nu}\partial_{\mu}\partial_{\nu} A(x)= J(x),
\ee
 as $\int d^{4}f\delta(f^2)f^{\mu}\partial_{\mu}\partial_{\tau}A_{f}(x)=0$ because \cite{hep-th/9708066} $A_{f}(x)=A_{-f}(x)$. \\ So, the standard formalism is retrieved from this $f$-formalism with the $A(x)$ as the average of $A_{f}(x)$, in the sense of (\ref{s}). We shall expose next the structure of Classical Electrodynamics written in terms of $A_{f}(x).$
\begin{center}
The Green's function
\end{center}

The $f$-wave equation (\ref{wef}) can be solved by a $f$-Green's function,
\be
\label{sgf}
A_{f}(x,\tau_{x})=\int d^{4}y\;d\tau_{y}\; G_{f}(x-y,\tau_{x}-\tau_{y})\;J(y),
\ee
with $G_{f}(x-y,\tau_{x}-\tau_{y})$ being a solution of
\be
\label{gfe}
\eta^{\mu\nu}\nabla_{\mu}\nabla_{\nu}G_{f}(x-y,\tau_{x}-\tau_{y})=\delta^{4}(x-y)\delta(\tau_{x}-\tau_{y}):=\delta^{5}(x-y).
\ee
This equation has been solved in \cite{hep-th/9708066}:
\be
\label{pr99}
G_{f}(x,\tau)=\frac{1}{2}\theta(-b{\bar f}.x)\theta(b\tau)\delta(\tau+f.x),
\ee
or
\be
\label{pr9}
G_{f}(x,\tau)=\frac{1}{2}\theta(bf^{4}t)\theta(b\tau)\delta(\tau+  f.x);
\ee
where $b =\pm1,$ and is related to the photon creation and annihilation processes.  For $f^{\mu}=({\vec f}, f^{4})$, ${\bar f}$ is defined by ${\bar f}^{\mu}=(-{\vec f}, f^{4}).$ $\theta (x)$ is\footnote{Writing $G_{f}(x,\tau)$, instead of $G_{f}(x-z,\tau_{x}-\tau_{z})$, implies that we have put our coordinate origins on the charge's retarded (advanced) position; in other words, we put $\tau_{s}=0$ and $z(\tau_{s}=0)=0$, where $s$ stands for retarded (advanced).} the step function, $\theta(x\ge0)=1$ and $\theta(x<0)=0.$  $G_{f}(x,\tau)$ does not depend on ${\vec x}_{\hbox {\tiny T}}$, where the subscript ${\hbox {\tiny T}}$ stands for transversality with respect to ${\vec f},$: ${\vec f}.{\vec x}_{{\hbox {\tiny T}}}=0.$  So, $d^4 y$ of eq. (\ref{sgf}) is effectively replaced by $dy_{\hbox {\tiny L}}d\tau_{y}dt_{y}$, with $y_{\hbox {\tiny L}}\equiv {\vec y}.\frac{\vec f}{|{\vec f}|}$: the support of $G_{f}(x,\tau)$ is reduced to the fibre $f$.\\ The properties of $G_{f}$ are discussed in \cite{hep-th/9708066}. $f$ and ${\bar f}$ are two opposing generators on the vertex of a same lightcone; they are associated, respectively, to the $b=+1$ and to the $b=-1$ solutions and, therefore, to the processes of creation and annihilation of a photon.
For $b=+1$, which implies on $t>0$, $G_{f}(x,\tau)$ describes a point signal emitted by the electron  at $\tau_{ret},$ and that has propagated to $x$ along the fibre $f,$ of the future ($f^4>0$) lightcone of $z(\tau_{ret})$;  for $b=-1$, which implies on  $t<0,$  $G_{f}(x,\tau)$ describes a point signal that is propagating  along the fibre $\bar{f}$ of the future lightcone of $x$ towards the point $z(\tau_{adv})$ where it will be absorbed (annihilated) by the electron. See the Figure 3.
\noindent Observe the differences from the standard interpretation of the Li\`enard-Wiechert solutions. There is no advanced, causality violating solution here. $J$ is the source of the $f$-solution and a sink for the $\bar{f}$-solution. These two solutions correspond to creation and annihilation of particles, of classical photons. 
\vglue2cm

\parbox[]{7.5cm}{
\begin{figure}
\vglue-10cm
\epsfxsize=400pt
\epsfbox{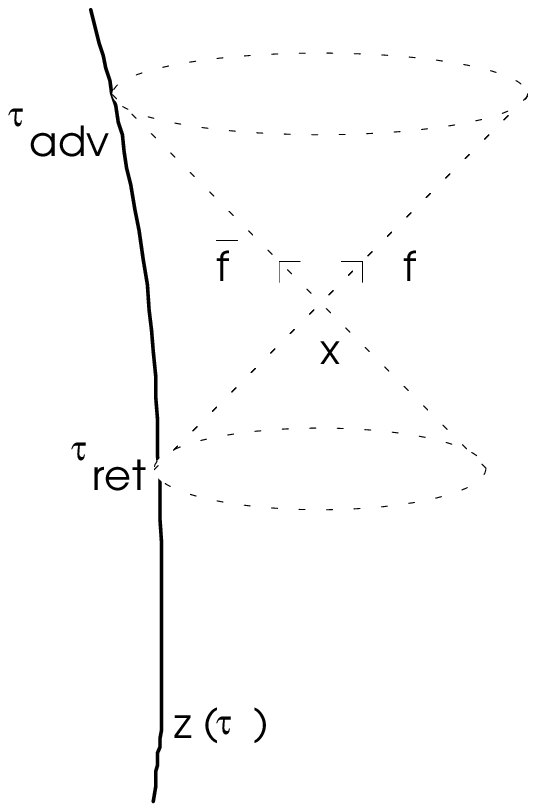}
\vglue-5cm
\end{figure}
\vglue-5cm}\\
\mbox{}
\hfill
\hspace{5.0cm}
\parbox[]{7.5cm}{\vglue-2.5cm Fig. 3. Creation and annihilation of classical particles as solutions of the wave equation. At the event $x$ there are two photons: one, on the fibre $f$, was created by the charge at $\tau_{ret}$, and the other one at the fibre $\bar{f}$  will be annihilated by the charge at $\tau_{adv}$. They are both retarded solutions. The charge is the source for the first particle and a sink for the second one.
}\\ \mbox{}
\vglue2cm

$G_{f}$ has no singularity, in contradistinction to the standard Green's function
\be
\l{sg} 
G(x,\tau=0)=2\delta(r^2-t^2),
\ee
 which, according to (\ref{s}), is retrieved \cite{hep-th/9708066} by 
\be
\l{gg}
G(x,\tau)=\frac{1}{2\pi}\int d^4f\delta(f^2)G(x,\tau)_{f}.
\ee
This process goes in the following way.  We write $f.x=f_{4}t+r|{\vec f}|\cos\theta_{f},$ where the angle $\theta_{f}$ defined by 
\be
\l{theta}
{\vec f}.{\vec x}\equiv r|{\vec f}|\cos \theta_{f},
\ee
for a fixed ${\vec x}$, which we take as our coordinate z-axis. Then we have, after the integration of $f_{4}$,
\be
\l{gg2}
G(x,\tau)=\frac{\theta(bf^{4}t)\theta(b\tau)}{8\pi}\int |{\vec f}| d|{\vec f}|d^2\Omega_{f}\{\delta{\big(}\tau+|{\vec f}|(r\;\cos\theta_{f}+t){\big)}+\delta{\big(}\tau+|{\vec f}|(r\;\cos\theta_{f}-t){\big)}\}
\ee
and then
$$
G(x,\tau)=-\frac{\theta(bf^{4}t)\theta(b\tau)}{8\pi}\int d^2\Omega_{f}\{\frac{\tau}{(r\;\cos\theta_{f}+t)|r\;\cos\theta_{f}+t|}+\frac{\tau}{(r\;\cos\theta_{f}-t)|r\;\cos\theta_{f}-t|}\}=
$$
$$
=-\frac{\theta(bf^{4}t)\theta(b\tau)}{4}\int_{-1}^{1}d\cos\theta_{f} \{\frac{\tau}{(r\;\cos\theta_{f}+t)|r\;\cos\theta_{f}+t|}+\frac{\tau}{(r\;\cos\theta_{f}-t)|r\;\cos\theta_{f}-t|}\}=\nonumber
$$
\be
\l{qq}
=\theta(bf^{4}t)\theta(b\tau)\frac{\tau}{t^2-r^2}.
\ee
So,
\be
\l{qqq}
G(x,\tau=0)=\cases{0,& for $t^2-r^2\ne0$;\cr
                   \infty,& for $t^2-r^2=0$.\cr}
\ee
We are treating $\tau$ and $x$ as independent variables; the constraint (\ref{1}) is used only afterwards as we have used it (as $|t|\ge r$) on the last step of (\ref{qq}), and on the second line of (\ref{qqq}) for resolving the $\frac{0}{0}$-indetermination with the L'H\^opital rule. Besides,
\be
\int_{0}^{\infty}G(x,\tau)d(\sqrt{t^2-r^2})=\theta(bf^{4}t)\theta(b\tau)\tau\int_{0}^{\infty}\frac{d(\sqrt{t^2-r^2})}{t^2-r^2}=\theta(bf^{4}t)\theta(b\tau)\frac{\tau}{\sqrt{t^2-r^2}}{\big|}_{t^2=r^2},
\ee
so that
\be
\int_{0}^{\infty}G(x,\tau=0)d(\sqrt{t^2-r^2})=1.
\ee
This justifies eqs. (\ref{sg}) and (\ref{gg}).

 \begin{center}
Physical and unphysical photons
\end{center}
There is more than just replaceing (\ref{sg}) in (\ref{pr9}) in order to get 
(\ref{gg}). For a real photon $A_{f}$ the equation (\ref{theta}) makes no sense because $f$ is the photon four-vector velocity, it is co-linear to $x$ and so $\theta_{f}=0.$ But the presence of $\cos\theta_{f}$ in (\ref{gg2}) is essential for getting (\ref{sg}). We can understand (\ref{theta}) and (\ref{gg2}) in the following way: we have substituted the single physical photon $A_{f}$ by a continuous and isotropic distribution of fictitious photons $A_{f'}$ that are characterized for not having $f'$ co-linear to their direction of propagation $x$ ($x$ and $\tau$ are fixed in both sides of equation (\ref{gg2})). As we will see in the following, these fictitious photons $A_{f'}$ do not satisfy the gauge condition and, consequently, are not transversal photons; they represent four degrees of freedom, not just two as $A_{f}$ does. So, in order to reproduce the usual continuous-field-based formalism we necessarily need to introduce the unphysical (scalar and longitudinal) photons. The continuous field $A(x,\tau)$ necessarily contains spurious, non-physical gauge-violating photons. This traces the borderline between $A_{f}(x,\tau)$ and $A(x,\tau)$ as also between their correspondent formalisms.\\
We can also see from (\ref{pr9}) that the classical photon $A_{f}$ doesn't present any singularity. The singularity presented by its average $A(x)$, as shown on its Green's function (\ref{gg}), is a consequence of this averaging process on its definition (\ref{gg}).\\

The virtual photons of the standard quantum field theory correspond to these unphysical photons. They could be used here, like they are used there in the standard quantum field theory, for a pictorical explanation of how the interactions between two a-distance-separated charges can interact trough the exchange of photons, real and virtual. The real photon, emitted by one charge, is the one that hits the other charge; the others, the virtual ones, do not contribute and must not be considered in the dynamics. This, of course, is just a verbal description; we'd rather stand with the position defined in \cite{hep-th/9610145}.
\section{The photon field}
Let us now apply this $f-$formalism to the electromagnetic field generated by a classical, spinless, point electron. In this formalism where $\tau$ is treated as an independent fifth parameter, a definition of a four-vector current must carry an additional constraint expressing the causal relationship between two events $y$ and $z$. Its four-vector current is given by
\be
\l{J}
J^{\mu}(y,\tau_{y})= eV^{\mu}(\tau_{z})\delta^{3}({\vec y}-{\vec z})\delta(\tau_{y}-\tau_{z}),
\ee
where $z^{\mu}(\tau_{z}),$ is the electron worldline parameterized by its proper-time $\tau_{z}.$ The sub-indices indicate their respective events. For $bf^{4}>0$, that is, for the field emitted by J, we have
\be
A^{\mu}_{f}(x,\tau)=2e\int d^{5}y G_{f}(x-y)V^{\mu}(\tau_{y})\delta^{3}({\vec y}-{\vec z(\tau)})\delta(\tau_{y}-\tau_{z}),
\ee
where the factor 2 accounts for a change of normalization with respect to (\ref{sgf}) as we are here excluding the annihilated photon. Then,
\be
\l{Af1}
A^{\mu}_{f}(x,\tau)=eV^{\mu}(\tau_{z})\theta[-{\bar f}.(x-z)]\theta[-f.(x-z)]{\Big |}_{d x.\Lambda^{f}.d x=0}.
\ee
So, the field $A_{f}$ is given, essentially, by the charge times its four-velocity at its retarded time. It is restricted by $\Delta\tau+f.(x-z)=0$ which is reduced to $f.(x-z)=0$ with the field massless condition $\Delta\tau=0$ and it means that the point $x,$ where the field is being observed, and the electron retarded position $z(\tau)$ must belong to a same lightcone generator $f.$  More informations can be extracted from this constraint as $\nabla_{\mu}f.(x-z)=0$ implies on $f_{\alpha}(\delta^{\alpha}_{\mu}+V^{\alpha}f_{\mu}){\Big |}_{d x.\Lambda^{f}.d x=0}=0$ or
\be
\l{fv} 
f.V{\Big |}_{d x.\Lambda^{f}.d x=0}=-1.
\ee
This relation may be seen as a covariant normalization of the time component of $f$ to 1 in the electron rest-frame at its retarded time, 
\be
\l{f4}
f^{4}{\Big |}_{{{\vec V}=0}\atop{\tau_{ret}}}=|{\vec f}|{\Big |}_{{{\vec V}=0}\atop{\tau_{ret}}}=1.
\ee
With a further derivation and with $a^{\mu}=\frac{dV^{\mu}}{d\tau}$ we get from (\ref{fv}) the following important relationship
\be
\l{dA0}
a.f{\Big |}_{d x.\Lambda^{f}.d x=0}=0,
\ee
between the direction $f$ of the photon emission (absorption) and the instantaneous change in the electron state of motion at the retarded (advanced) time. The fictitious photons $A_{f'}$ introduced in (\ref{gg2}) does not satisfy (\ref{dA0}) as $\nabla_{\mu}f'.(x-z)=0 $ implies on $f'_{\alpha}(\delta^{\alpha}_{\mu}+V^{\alpha}f_{\mu}){\Big |}_{d x.\Lambda^{f}.d x=0}=0$ or
\be
\l{f'}
f'_{\mu}+f_{\mu}f'.V{\Big |}_{d x.\Lambda^{f}.d x=0}=0
\ee
and this equation does not make any sense unless $f'\equiv f,$ as can be easily seen by contracting (\ref{f'}) with $f^{\mu},\;f'^{\mu},\;V^{\mu}$ and $a^{\mu}$.  It means that $A_{f'}$ does not satisfy the condition (\ref{dA0}) and this has an important physical significance, as we will see next.\\
The Maxwell field $F_{f}$ at the fibre $f$ is defined to be  $F^{f}_{\mu\nu}=(\nabla_{\mu}A^{f}_{\nu}-\nabla_{\nu}A^{f}_{\mu})$, but let us show now that $\nabla\theta[f.(x-z)]$ and $\nabla\theta[{\bar f}.(x-z)]$ do not contribute to $\nabla A_{f},$ except at $x=z(\tau),$ as a further consequence of the field constraint.  
The derivation of $\theta[{\bar f}.(x-z)]$ generates a $\delta[{\bar f}.(x-z)]$ which, with $f.(x-z)=0$, requires $x=z$, and $$\nabla_{\mu}\theta[f.(x-z)]=\delta[f.(x-z)]f_{\mu}(1+f.V)=0$$ because of (\ref{fv}).\\
For notation simplicity (see the footnote number 2) we take, from now on, $\tau_{ret}=0$ and $z(\tau_{ret}=0)=0$, so that we can replace $\Delta\tau$ and $\Delta x$ by $\tau$ and $x$, respectively. So, for $x\ne0$ and $t>0$ we write just

\be
\l{dA}
\nabla_{\nu}A^{\mu}_{f}=\nabla_{\nu}(eV^{\mu})=-ef_{\nu}a^{\mu}{\Big |}_{d x.\Lambda^{f}.d x=0}.
\ee
The Lorentz gauge condition $\nabla.A_{f}=0,$  which is necessary for reducing $\nabla_{\nu}F_{f}^{\mu\nu}$ to $\nabla_{\nu}\nabla^{\nu}A_{f}^{\mu}$, implies on (\ref{dA0}) and so it is automatically satisfied by the field $A_{f}$ of (\ref{Af1}), but not by the fictitious fields $A_{f'}$. Thus, we have
\be
\l{FedA}
F^{f}_{\mu\nu}=-e(f_{\mu}a_{\nu}-f_{\nu}a_{\mu}){\Big |}_{d x.\Lambda^{f}.d x=0}
\ee
that describes the electromagnetic field of a classical photon $A_{f}$:
\be
\l{Ef}
E_{f}^{i}=-F^{4i}_{f}{\Big |}_{d x.\Lambda^{f}.d x=0}=e(f^{4}a^{i}-f^{i}a^{4}){\Big |}_{d x.\Lambda^{f}.d x=0},
\ee
and 
\be
\l{Bf}
B^{i}_{f}=-\epsilon_{ijk}F^{jk}_{f}{\Big |}_{d x.\Lambda^{f}.d x=0}=e\epsilon_{ijk}f^{j}a^{k}{\Big |}_{d x.\Lambda^{f}.d x=0}.
\ee
Then we see  that ${\vec E}_{f},\;{\vec a}$ and ${\vec f}$ belong to a same plane, that ${\vec B}_{f}$ is by definition orthogonal to ${\vec f}$, and therefore, that ${\vec E}_{f}$ and ${\vec B}_{f}$ are orthogonal to each other
$${\vec f}.{\vec B}_{f}=0,$$
$${\vec E}_{f}.{\vec B}_{f}=0,$$
and that, as far as the solution (\ref{Af1}) automatically satisfies the gauge condition (\ref{dA0}), there is no unphysical (scalar and longitudinal) photon i.e.  solutions with longitudinal electric field (${\vec E}_{f}=ef^4{\vec a}_{\hbox {\tiny T}}$):
$${\vec f}.{\vec E}_{f}=e(f^{4}{\vec a}.{\vec f}-a^{4}{\vec f}.{\vec f}){\Big |}_{d x.\Lambda^{f}.d x=0}=ef^{4}a.f{\Big |}_{d x.\Lambda^{f}.d x=0}=0.$$  ${\vec E}_{f},\;{\vec B}_{f}$ and ${\vec f}$ form a triad of orthogonal vectors.\\
The $f-$Poynting vector ${\vec S}{_f}$ is defined by (without the standard $\frac{1}{4\pi}$ factor, which is associated to its average ${\vec S}$)
\be
\l{pv}
{\vec S}_{f}={\vec E}_{f}\times{\vec B}_{f}= e^{2}\{{\vec f}(f^{4}{\vec a}.{\vec a}-a^{4}{\vec a}.{\vec f})-{\vec a}(f^{4}{\vec a}.{\vec f}-{\vec f}.{\vec f}a^{4})\}{\Big |}_{d x.\Lambda^{f}.d x=0},
\ee
which, with (\ref{dA0}), is reduced to  
\be
\l{PV}
{\vec S}_{f}={\vec E}_{f}\times {\vec B}_{f}={\vec f}e^{2}f^{4}a^{2} {\Big |}_{d x.\Lambda^{f}.d x=0},
\ee
and this confirms the physical interpretation of $A_{f}$ as a classical photon propagating along the fibre $f$. 

\section{The photon angular distribution}

The Lorentz condition (\ref{dA0}) restricts the possible direction of emission of a classical photon by an accelerated charge: ${\vec f}$ is orthogonal to ${\vec a}$ in the charge instantaneous rest-frame:
\be
\l{af}
{\vec a}.{\vec f}{\Big |}_{{{\vec V}=0}\atop{d x.\Lambda^{f}.d x=0}}=0,
\ee
as $a.V=0$ requires $a^{4}{\Big |}_{{{\vec V}=0}\atop{d x.\Lambda^{f}.d x=0}}=0.$ \\

In laboratory  $\vec{a}$ is known only up to an experimental uncertainty $\Delta \vec{a}$. See the Figure 4.

\vspace{-3.0cm}

\hspace{-4.0cm}
\parbox[]{7.5cm}{\hspace{-2cm}
\begin{figure}
\vglue-1.5cm
\hspace{1.5cm}
\epsfxsize=400pt
\epsfbox{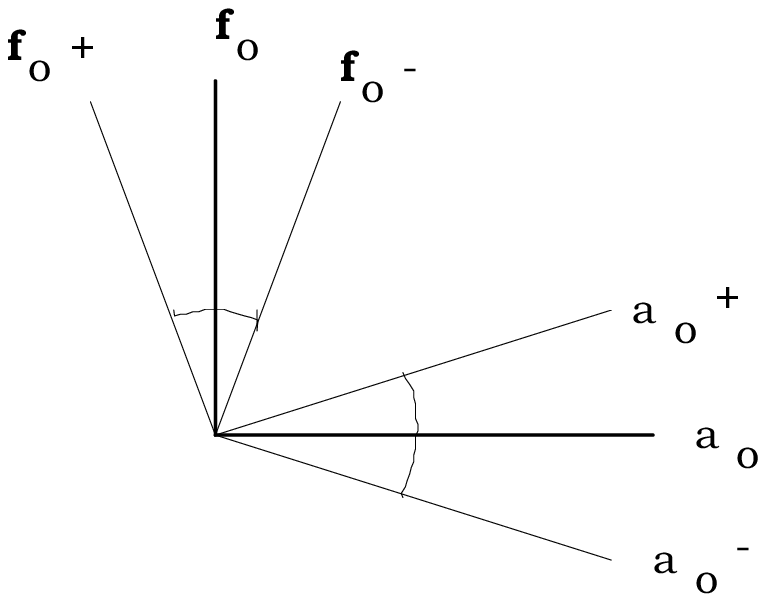}
\vglue-6cm
\end{figure}
\vglue-8cm
}
{}\\
\mbox{}
\vglue1cm
\hfill
\hspace{6.0cm}
\parbox[]{7.5cm}{Fig. 4. $\vec{a}$ is known up to an experimental uncertainty $\Delta \vec{a}$.  Then there corresponds an indetermination $\Delta {\vec f}$, such that whichever be $\vec{a}$, inside its range $[\vec{a}_{0}-\Delta\vec{a},\vec{a_{0}}+\Delta\vec{a}],$ ${\vec f}$, in its range $[\vec{f}_{0}-\Delta\vec{f},\vec{f_{0}}+\Delta\vec{f}],$ will be in the plane orthogonal to $\vec{a}.$ }\\
 \mbox{}

\vglue1.5cm
Then there corresponds, according to (\ref{af}), an indetermination $\Delta {\vec f}$, such that whichever be $\vec{a}$, inside its range $[\vec{a}_{0}-\Delta\vec{a},\vec{a_{0}}+\Delta\vec{a}],$ ${\vec f}$, in its range $[\vec{f}_{0}-\Delta\vec{f},\vec{f_{0}}+\Delta\vec{f}],$ will be in the plane orthogonal to $\vec{a}$, and that
\be
\vec{a}_{0}.\Delta\vec{f}+\vec{f}_{0}.\Delta\vec{a}=0,
\ee
\be
\vec{a}_{0}.\vec{f}_{0}=0,
\ee
are satisfied.
$\vec{a}_{0}$ and $\vec{f}_{0}$ denote the most probable value of, respectively,  $\vec{a}$ and $\vec{f}$.\\

The photon polarization, according to (\ref{Ef}) is determined by the charge instantaneous acceleration.
If the electron motion is such that ${\vec V}$ and ${\vec a}$ are collinear vectors, then with a boost along ${\vec V}$, we have for the angle $\theta$ between ${\vec V}$ and ${\vec f}_{0}$:
$$\tan\theta=\frac{\sin\theta'}{\gamma(\beta+\cos\theta')}=\frac{1}{\gamma\beta},$$ or $$\sin\theta=\sqrt{1-\beta^{2}},$$ where $\beta= |\frac{d{\vec z}}{dt_{z}}|$ and $\gamma=(1-\beta^{2})^{-\frac12}.$ So, a ultra-relativistic electron emits a photon in a very narrow cone about the ${\vec V}$ direction. The direction of photon emission coincides with the direction of maximum emission in the standard formalism \cite{Jackson,Ternov}. The photon polarization again, as we see from (\ref{Ef}), is along the electron instantaneous acceleration, regardless the electron velocity. See the Figure 5, where ${\vec f}'$ and ${\vec a}'$ are, respectively, ${\vec f}$ and ${\vec a}$ on the electron instantaneous rest frame.
If the charge is in a circular motion then ${\vec a}.{\vec V}=0$, which from  $a.V=0$ implies also that $a^{4}=0$ and then, with (\ref{dA0}), that ${\vec a}.{\vec f}=0.$ Inverting (\ref{Bf}) we find that
\be
\l{fv1}
{\vec f}=\frac{{\vec a}\times{\vec B}_{f}+e{\vec a}{\vec f}.{\vec a}}{e{\vec a}.{\vec a}}=\frac{{\vec a}\times{\vec B}_{f}}{e{\vec a}.{\vec a}}.
\ee
\vspace{-5.0cm}

\hspace{-4.0cm}
\parbox[]{7.5cm}{\hspace{-2cm}
\begin{figure}
\vglue-1.5cm
\hspace{1.0cm}
\epsfxsize=400pt
\epsfbox{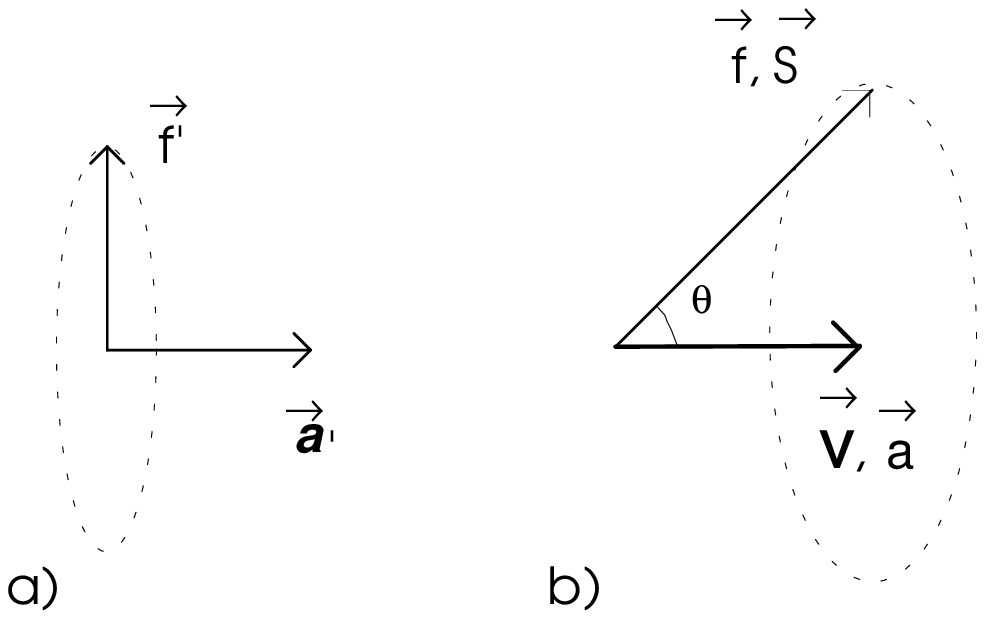}
\vglue-6cm
\end{figure}
\vglue-8cm
}
{}\\
\mbox{}
\vglue1cm
\hfill
\hspace{6.0cm}
\parbox[]{7.5cm}{Fig. 5. Radiation from an electron in a straight-line motion. The Lorentz condition requires that ${\vec f}'$ be orthogonal to the electron three-vector acceleration ${\vec a}'$ on its instantaneous rest frame: ${\vec a}'.{\vec f}'=0.$ In the laboratory frame the radiation is restricted to a cone of aperture $\theta$, with $sin\theta=\sqrt{1-\beta^{2}},$ $\quad{\vec \beta}=\frac{d{\vec x}}{dt}$.}\\
 \mbox{}

\vglue1.5cm
 
\noindent Then we conclude from (\ref{fv1}) that ${\vec f}$, for a charge in a circular motion,  is orthogonal to the plane defined by ${\vec B}_{f}$  and ${\vec a}.$ But, in the absence of magnetic monopoles, ${\vec B}_{f}$ is always orthogonal to ${\vec V}$ and so we have that the field synchrotron radiation is always emitted in the ${\vec V}$-direction. Again, this agrees with the direction of maximum emission in the standard formalism \cite{Jackson,Ternov}.

\section{The photon energy-momentum}

The invariant $F_{f}^{2}=F_{f}^{\mu\nu}F^{f}_{\mu\nu}$  takes the following form
\be
\l{F2}
F_{f}^{2}=-2e^{2}(a.f)^{2}{\Big |}_{\tau+f.x=0}=0,
\ee
and therefore, $|{\vec E}_{f}|=|{\vec B}_{f}|,$ as a consequence of (\ref{dA0}).
The electron self-field energy-momentum tensor at the fibre $f,$ defined by  (without the standard $\frac{1}{4\pi}$ factor)
$$\Theta_{f}^{\mu\nu}=F_{f}^{\mu\alpha}\eta_{\alpha\beta}F_{f}^{\beta\nu}-\frac{\eta^{\mu\nu}}{4}F_{f}^{2}$$
becomes
\be
\l{t}
\Theta_{f}^{\mu\nu}=e^{2}{\Big\{}a_{f}(a^{\mu}f^{\nu}+a^{\nu}f^{\mu})-f^{\mu}f^{\nu}a^{2}+ \frac{\eta^{\mu\nu}}{2}a_{f}^{2}{\Big\}}{\Big |}_{d x.\Lambda^{f}.d x=0},
\ee
where $a_{f}=a.f,$
and with (\ref{dA0}) it is reduced to
\be
\l{tr}
\Theta_{f}^{\mu\nu}=-e^{2}f^{\mu}f^{\nu}a^{2}{\Big |}_{d x.\Lambda^{f}.d x=0}.
\ee
Its physical interpretation is quite clear: $\Theta_{f}$ has support only on the fibre $f$, it is not null only at those events that satisfy the constraint $\tau+  f.x=0$ on the fibre $f$ that passes by the charge at the emission (retarded) time. Clearly it represents the energy-momentum of a point ``photon" propagating along the lightcone generator $f$.\\
In the standard formalism the electromagnetic momentum-energy four-vector P is the flux of $\Theta$ through a hypersurface $\sigma$,
\be
\l{Ps}
P^{\mu}=-\int_{\sigma} d\sigma^{3}\Theta^{\mu\nu}n_{\nu},
\ee
where $n_{\nu}$ is a vector normal to the hypersurface $\sigma.$ In the case of extended causality with $\tau$ treated as fifth independent parameter, the expression of $P_{f}$, as the support of $\Theta_{f}$ is the fibre $f$, is reduced to
\be
\l{P}
P_{f}^{\mu}=-\int_{\sigma} d^{4}\sigma\Theta_{f}^{\mu\nu}n_{\nu}= -\Theta^{\mu\nu}_{f}n_{\nu}{\Big |}_{\tau+f.x=0}=e^{2}a^{2}f^{\mu}{\Big |}_{\tau+f.x=0},
\ee
with a space-like n, defined by $n=f-V$, which, in the charge rest-frame, reduces to $n{\bigg|}_{{\vec V}=0}=(0,{\hat n})$, where ${\hat n}$ is the normal of a spherical surface centred at the coordinate origin (the charge position)\cite{Teitelboim}; $f.n=1.$
and $  n.V=0$.
While (\ref{Ps}) is meaningful \cite{hep-th/9610028} only for $r>0$ in order to avoid the singularity of (\ref{sg}) at $r=0$, $P_{f}$ is valid without restrictions.
It is important to underline that $P_{f}$ is not null only at a single point, the instantaneous photon position at its fibre $f.$ This agrees with our interpretation of $A_{f}$ as a photon on a fibre $f.$
In a theory of discrete classical fields $P_{f}^{\mu}=e^{2}a^{2}f^{\mu}$ is the energy-momentum  four-vector radiated by the charge $e$ under the acceleration $a$ at the instant of time $\tau=\tau_{ret}$ through the emission of a single photon $A_{f}(x,\tau)$ along the fibre $f$. See the Figure 6. It can be written as 
\be
\l{pff}
P^{\mu}_{f}=e^{2}\int d\tau a^{2}f^{\mu}\delta(\tau-\tau_{ret}),
\ee
or as 
\be
\l{pff1}
\frac{d P^{\mu}_{f}}{d \tau}=e^{2}a^{2}f^{\mu}\delta(\tau-\tau_{ret}),
\ee
\vspace{-5.0cm}

\hspace{-3cm}
\parbox[]{7.5cm}{\hspace{-2cm}
\begin{figure}
\vglue-3cm
\epsfxsize=400pt
\epsfbox{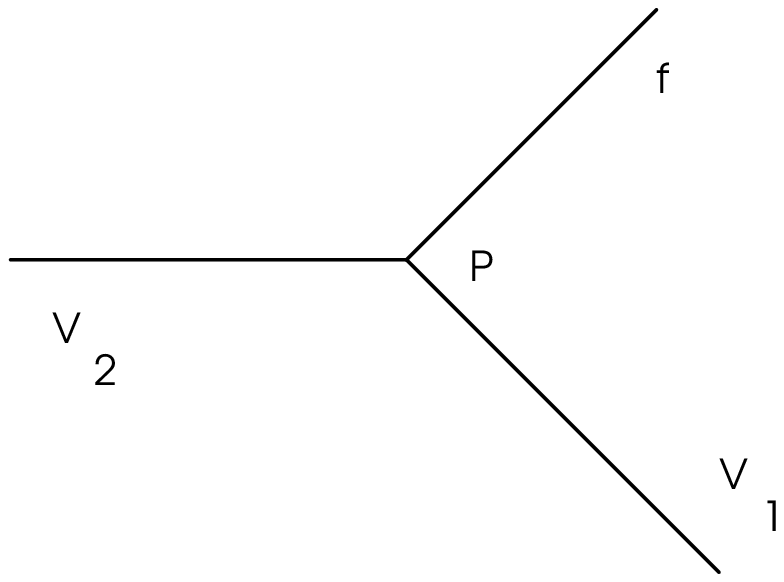}
\vglue-6cm
\end{figure}
\vglue-8cm
}
{}\\
\mbox{}
\vglue3cm
\hfill
\hspace{6.0cm}
\parbox[]{7.5cm}{Fig. 6. In the standard formalism the radiation emission (or absorption) is a continuous process but in the $f-$formalism it is discrete; it just occurs at isolated events, like P on the figure, on the charge worldline $V_{1}V_{2}$.}\\ \mbox{}
\vglue.5cm 

This expression is very distinct from the well-known Larmor Theorem which is derived with the use of a continuous and distributed field $A(x,\tau)$. In order to retrieve it we must again replace the single photon field $A_{f}$ by a fictitious continuous distribution of $A_{f'}$, which, as we know, do not obey (\ref{dA0}). Then instead of (\ref{pff}) we must use (\ref{t}) for getting 
\be
\l{P'}
P_{f'}^{\mu}=\int d\tau \Theta^{\mu\nu}{\big|}_{\tau_{ret}}n_{\nu}=e^{2}\{(a_f'^{2}-a^{2})f'^{\mu}+a^{\mu}a_{f'}+\frac{n^{\mu}}{2}a_{f'}^{2}\}{\Big |}_{\tau_{ret}}.
\ee
In the standard description of a continuous emission of a continuous and distributed field $A(x,\tau)$, the discrete and instantaneously emitted $P_{f}^{\mu}$ is replaced by its average (observe the return of the $\frac{1}{4\pi}$ factor).
\be
\l{Pf'}
 P^{\mu}=\frac{1}{4\pi}\int d^{2}\Omega_{f'}P_{f'}^{\mu},
\ee
where $d^{2}\Omega_{f'}=d\phi_{f'}sin\theta_{f'}\;d\theta_{f'}$ and $\phi_{f'}$ and $\theta_{f'}$ define a generic fibre $f'(\theta_{f'},\phi_{f'})$.

Now, using the following well known identities \cite{Teitelboim1,Synge}
$$\frac{1}{4\pi}\int d^{2}\Omega f'^{\alpha}=V^{\alpha},$$
$$\frac{1}{4\pi}\int d^{2}\Omega f'^{\alpha}f'^{\beta}=\frac{\Delta^{\alpha\beta}}{3}+V^{\alpha}V^{\beta},$$
$$\frac{1}{4\pi}\int d^{2}\Omega f'^{\alpha}f'^{\beta}f^{\gamma}=\Delta^{(\alpha\beta}V^{\gamma)}+V^{\alpha}V^{\beta}V^{\gamma},$$
where $\Delta=\eta+VV$, and the parenthesis on the superscripts means total symmetrization, we find from (\ref{Pf'}) and (\ref{P'}) that
\be
\l{BP}
\frac{dP^{\mu}}{d\tau}=\frac{2}{3}e^{2}a^{2}V^{\mu},
\ee
the standard expression of the Larmor Theorem.
The momentum-energy $e^{2}a^{2}f^{\mu}{\Big |}_{d x.\Lambda^{f}.d x=0}$ of a single photon is then replaced in the continuum picture by its spacetime-smeared value (\ref{BP}).\\ 

As a final remark we observe that when the effective average field $A(x,\tau)$ is quantized in a manifestly covariant way we have problems with non-physical photons that, in contradistinction to the real ones, must be eliminated by the imposition of the Lorentz gauge $\partial.A=0$ as a constraint on the allowed  physical states \cite{Gupta}. This is so closely connected to our alternative approach in terms of $A_{f}$ fields (without the non-physical $A_{f'}$) that we could not avoid anticipating this comment on the quantization of the electromagnetic field, a subject to be properly discussed in a subsequent work; it justifies our conviction that one should quantize $A_{f}$ and not its average $A.$
\begin{center}
Acknowledgements
\end{center}
J.V. Costa and A. S. Pedra acknowledge their grants from CAPES for writing their respective M.Sc. dissertations. The authors thank Marcelo Sarandy for a critical reading of the manuscript.

\end{document}